\documentstyle[12pt,epsf]{article}

\begin{document}

\title{\bf Splitting solitons on a torus}
\author{R. J. Cova \\
{\em Department of Mathematical Sciences, University of Durham}, \\
{\em Durham DH1 3LE, England} \\
{\em Departamento de F\'{\i}sica FEC, Universidad del Zulia}, \\
{\em Apartado 526, Maracaibo, Venezuela}}

\maketitle

\abstract
{New $CP^1$-soliton behaviour on a flat torus is reported.
Defined by the Weierstrass elliptic function and numerically-evolved from
rest, each soliton splits up in two lumps which eventually reunite,
divide and get back together again, {\em etc.}. 
This result opens up the question of fractional topological charge.}

\section{Introduction \label{sec:intro}}

The $CP^1$ model in (2+1) dimensions appears as a low dimensional
analogue of non-abelian gauge field theories in four dimensional
space-time.  This analogy relies on common properties like conformal
invariance, existence of topological solitons, hidden symmetry and
asymptotic freedom. Amongst various applications, $CP^1$ models have
been used in the study of the quantum Hall effect and high-$T_{c}$
superconductivity. In differential geometry, the soliton-solutions of
$CP^1$ models are known as harmonic maps, a rich industry of research
on its own. 

The classical (2+0)-dimensional $CP^1$ or non-linear $O(3)$ model on
the extended plane \( \Re_{2} \cup \{\infty\} \approx S_2 \), where
the soliton solutions are harmonic maps \( S_2 \mapsto S_2 \), has
been amply discussed in the literature \cite{lit1,lit2}. In (2+1)
dimensions the model is not integrable, and the study of its dynamics
is done with the aid of numerical simulations.  Due to the conformal
invariance of the theory on the plane, the $O(3)$ solitons are
unstable in the sense that they change their size under any small
perturbation, either explicit or introduced by the discretisation
procedure. It can make the solitons shrink indefinitely and, when
their width is comparable to the lattice spacing, the numerical code
breaks down \cite{leese1}. However, such instability can be cured by
the addition of two extra terms to the lagragian \cite{leese2}. The
first one resembles the term introduced by Skyrme in his nuclear model
in four dimensional space-time \cite{sky}, and the second one is a
potential term. The fields of the planar Skyrme model (skyrmions)
produce stable lumps which repel each other when started off from rest
\cite{leese2,pms1}.

In a recent paper \cite{non} we considered both the pure and modified
$CP^1$ schemes imposing periodic boundary conditions, which amounts to
defining the system on a torus $T_2$. The corresponding soliton
configurations are harmonic maps \( T_2 \mapsto S_2 \). In \cite{non}
we found (using the Weierstrass' $\sigma(z)$ function to define the
solitons) that in contradistinction with the familiar theory on $S_2$,
the toroidal model: $\bullet$ has no {\em analytical} single-soliton
solution [this is because elliptic functions, in terms of which the
toroidal solitons must be expressed, are at least of the second order;
or, in the language of differential geometry, because genus(torus)=1];
$\bullet$ needs only a Skyrme term to stabilise the solitons (thus the
lagrangian retains its $O(3)$ invariance: on $S_2$, the latter is
broken by the additional potential term); $\bullet$ does not require a
damping set-up for the numerical simulation (a radiation-absorbing
device is implemented for the model evolved on the compactified plane
in order to prevent the reflection of kinetic waves from the
boundaries); $\bullet$ has perfectly static skyrmions when their
initial speed $v_0$ is zero (as already pointed out, on $S_2$ they
move away from each other for $v_0=0$); $\bullet$ possesses no
critical velocity below which the skyrmions scatter back-to-back in
head-on collisions. They always scatter at right angles provided $v_0
\ne 0$. Also, on $T_2$ the skyrmions scatter any number of times
(multi-scattering), as they keep encountering each other in the
periodic grid. 

In the present work we continue the study of periodic $CP^1$
configurations, limiting ourselves to those with $v_0$=0 in the topological
charge-two sector. Defining the solitons through the elliptic function
$\wp(z)$ of Weierstrass we will see new soliton
behaviour, where each skyrmion-lump divides itself in two smaller
components that glue back together, split and reunite again and so on. 

\section{Periodic skyrmion model} 

Our model is given by the lagrangian density
\begin{equation}
{\cal L}=
\frac{|\partial_t W|^{2}-2 |\partial_z W|^{2}}{(1+|W|^{2})^{2}} 
+8\theta_{1}\frac{|\partial_z W|^{2}}{(1+|W|^{2})^{4}}
     (|\partial_{t} W|^{2}-|\partial _{z} W|^{2}),
\label{eq:lagrangian}
\end{equation}
\( z=x+iy \; \epsilon \; T_2 \), which is the pure
$CP^1$ model plus an additional Skyrme, $\theta_1$-term
\( (\theta_1 \; \epsilon \; \Re^+) \).
The complex field $W$ obeys the periodic boundary condition
\begin{equation}
W[z+(m+in)L]=W(z), \qquad \forall t,
\label{eq:boundary}
\end{equation}
where $m,n=0,1,2,...$ and  $L$ is the size of a square torus. The static
solitons (skyrmions) are elliptic functions which may be written as
\begin{equation}
W=\lambda \, \wp(z-a) + b, 
\qquad \lambda, a, b \; \epsilon \;{\cal Z},
\label{eq:w}
\end{equation}
$\wp(z)$ being the elliptic function of Weierstrass. Within a 
fundamental cell of length $L$, $\wp$ possesses the expansion
\cite{goursat}
\begin{equation}
\wp(z)=z^{-2} + \xi_2 z^2 + \xi_3 z^4 + ... + \xi_j z^{2j-2} + ...,
\qquad \xi_j \; \epsilon \; \Re.
\label{eq:p}
\end{equation}
This function is of the second order, hence (\ref{eq:w}) represents
solitons of topological index 2. Note that (\ref{eq:w})
is an approximate solution of the model (\ref{eq:lagrangian}), except in the
pure $CP^1$ limit ($\theta_1$=0) where it exactly solves the corresponding
static field equation. Therefore, we expect our solitons to evolve only
for a non-zero Skyrme parameter.

In reference \cite{non} we computed the periodic solitons through
\begin{equation}
W=\prod_{j=1}^{\kappa} \frac{\sigma(z-a_{j})}{\sigma(z-b_{j})},
\quad \sum_{j=1}^{\kappa} a_{j}=\sum_{j=1}^{\kappa} b_{j},
\label{eq:wsigma}
\end{equation}
employing a subroutine that numerically calculates $\sigma(z)$. 
Via the formula below \cite{goursat}, in this paper we use the same 
subroutine to compute $\wp(z)$ :
\begin{equation}
\wp(z)=-\frac{d^2}{dz^2}\ln[\sigma(z)],
\label{eq:psig}
\end{equation}
where the Laurent expansion for $\sigma$ reads
\begin{equation}
\sigma(z)=\sum_{j=0}^{\infty}{c}_{j}z^{4j+1}, 
\qquad c_j \; \epsilon \; \Re.
\label{eq:sig}
\end{equation}

\section{Basic numerical procedure}

We treat configurations of the form (\ref{eq:w}) as the 
initial conditions for our time evolution, studied numerically. 
Our simulations run in the $\phi$-formulation of the
model, whose field equation follows from the lagrangian density 
(\ref{eq:lagrangian}) with the help of the stereographic projection
\begin{equation} 
W=\frac{\phi_{1}+i\phi_{2}}{1-\phi_{3}},
\label{eq:wphi}
\end{equation}
where the real scalar field \( \vec{\phi}=(\phi_1,\phi_2,\phi_3) \)
satisfies \( \vec{\phi}.\vec{\phi}=1 \). 

We compute the series (\ref{eq:sig}) up to the fifth term, the
coefficients $c_j$ being in our case negligibly small for $j \geq 6$. We
employ the fourth-order Runge-Kutta method and approximate the spatial
derivatives by finite differences. The laplacian is evaluated using the
standard nine-point formula and, to further check our results, a 13-point
recipe is also utilised.  Our results showed unsensitiveness to either
method. The discrete model evolves on a 200 $\times$ 200 periodic lattice
($n_x=n_y=200$) with spatial and time steps $\delta x$=$\delta y$=0.02 and
$\delta t$=0.005, respectively. The size of our fundamental, toroidal
network is $L=n_x \times \delta x=4$. 

Unavoidable round-off errors gradually shift the fields away from the
constraint \( \vec{\phi}.\vec{\phi}=1 \). So we rescale \( \vec{\phi}
\rightarrow \vec{\phi}/\sqrt{\vec{\phi}.\vec{\phi}} \) every few iterations. 
Each time, just before the rescaling operation, we evaluate the quantity \(
\mu \equiv \vec{\phi}.\vec{\phi} - 1 \)\, at each lattice point. Treating the
maximum of the absolute value of $\mu$ as a measure of the numerical errors,
we find that max$|\mu|$ $\approx$ 10$^{-8}$.  This magnitude is useful as a
guide to determine how reliable a given numerical result is. Usage of an
unsound numerical procedure in the Runge-Kutta evolution shows itself as a
rapid growth of max$|\mu|$; this also occurs, for instance, in the $O(3)$
limit ($\theta_1=0$) when the unstable lumps of energy become infinitely spiky.

\section{Splitting lumps}

\begin{figure}
\epsfverbosetrue
\centerline{\epsfbox{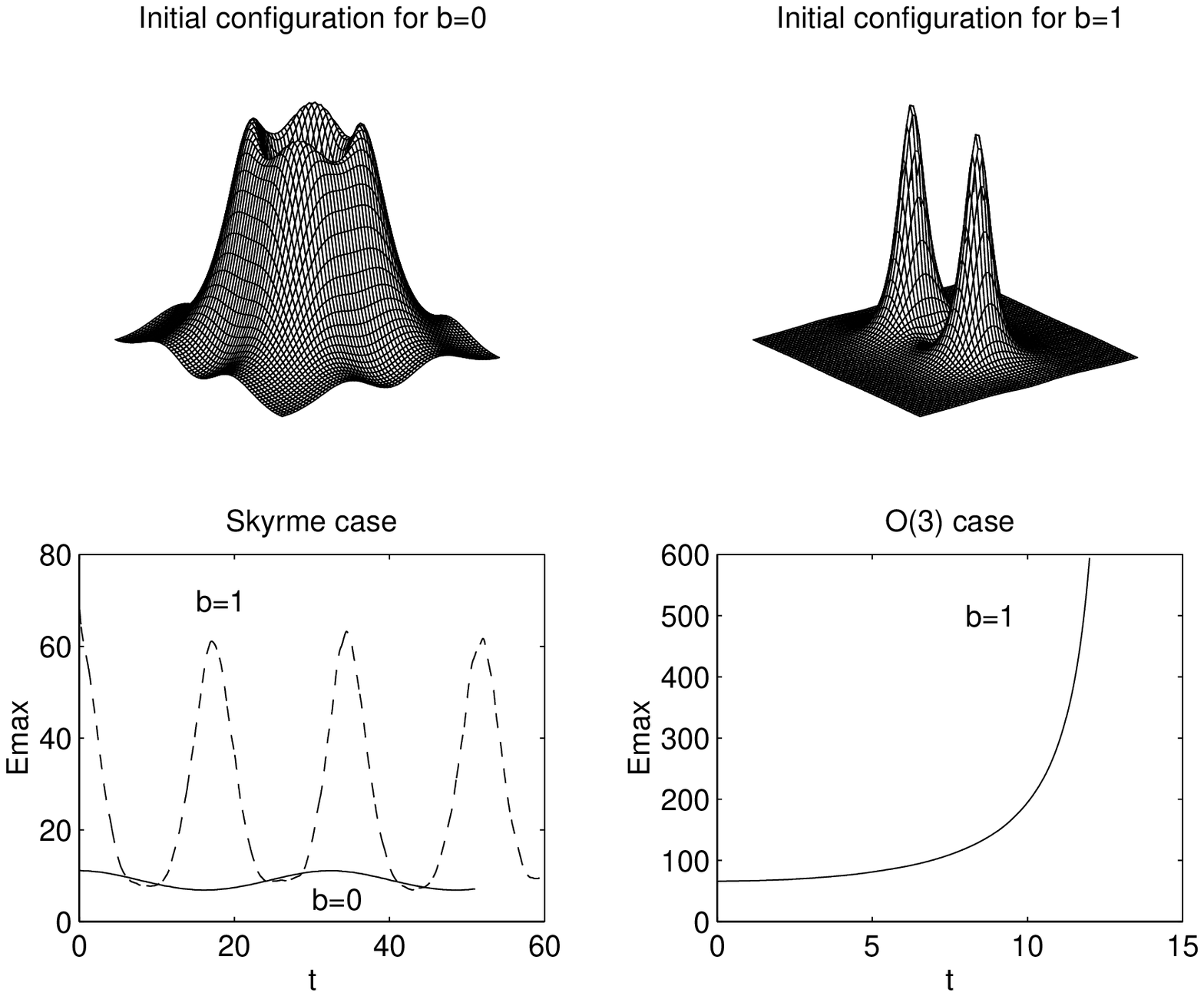}}
\caption{Energy density configurations at $t=0$ and the
evolution of their peaks. The bottom-right
graph corresponds to $\theta_1$=0, when the lumps are unstable and
shrink non-stoppingly.}
\label{fig:pfone}
\end{figure}

The energy density associated with our degree-2 solitons\linebreak \(
W= \lambda \wp(z-a)+b \) can be read-off from the lagrangian
(\ref{eq:lagrangian}). The pole $a$ determines the position of the
energy humps on the basic cell;  we shall take $a$=(2.025, 2.05).  The
quantity $b$ defines the distance between the solitons; for $b$=0 we
have \( W=\wp(z-a) \) (use $\lambda$=1 throughout, for simplicity),
whose energy density gives indistinguishable lumps on top of each
other, as depicted in figure \ref{fig:pfone} (top-left). The value
$b$=1 gives two lumps separated along the ordinates [figure
\ref{fig:pfone} (top-right)], whereas $b$=-1 positions them along the
abscissas. A pure imaginary $b$ places our extended structures on a
diagonal bisecting the toroidal grid [for $a$=(2,2) the bisection is
exact, but this is not a good numerical value because $z$, with
spatial steps of 0.02, blows $W$ at $z=a$], and $b$ with non-zero real
and imaginary parts situates the lumps in an arbitrary diagonal of the
cell. These set-ups are true regardless of $\theta_1$, which we have
put equal to 0.001. 

Our numerical simulations show that the skyrmions are stable. Their
stability is reflected on the left-hand side of the nether
half of figure \ref{fig:pfone}, which shows the evolution of the the maximum
value of the system's  total energy density ($E_{max}$) for $b$=0,1. In the
$O(3)$ limit the lumps are no longer stable, as can be appreciated from the
bottom-right graph of figure \ref{fig:pfone}. In this case, as expected, the
solitons remain static with the passing of time.

\begin{figure}
\epsfverbosetrue
\centerline{\epsfbox{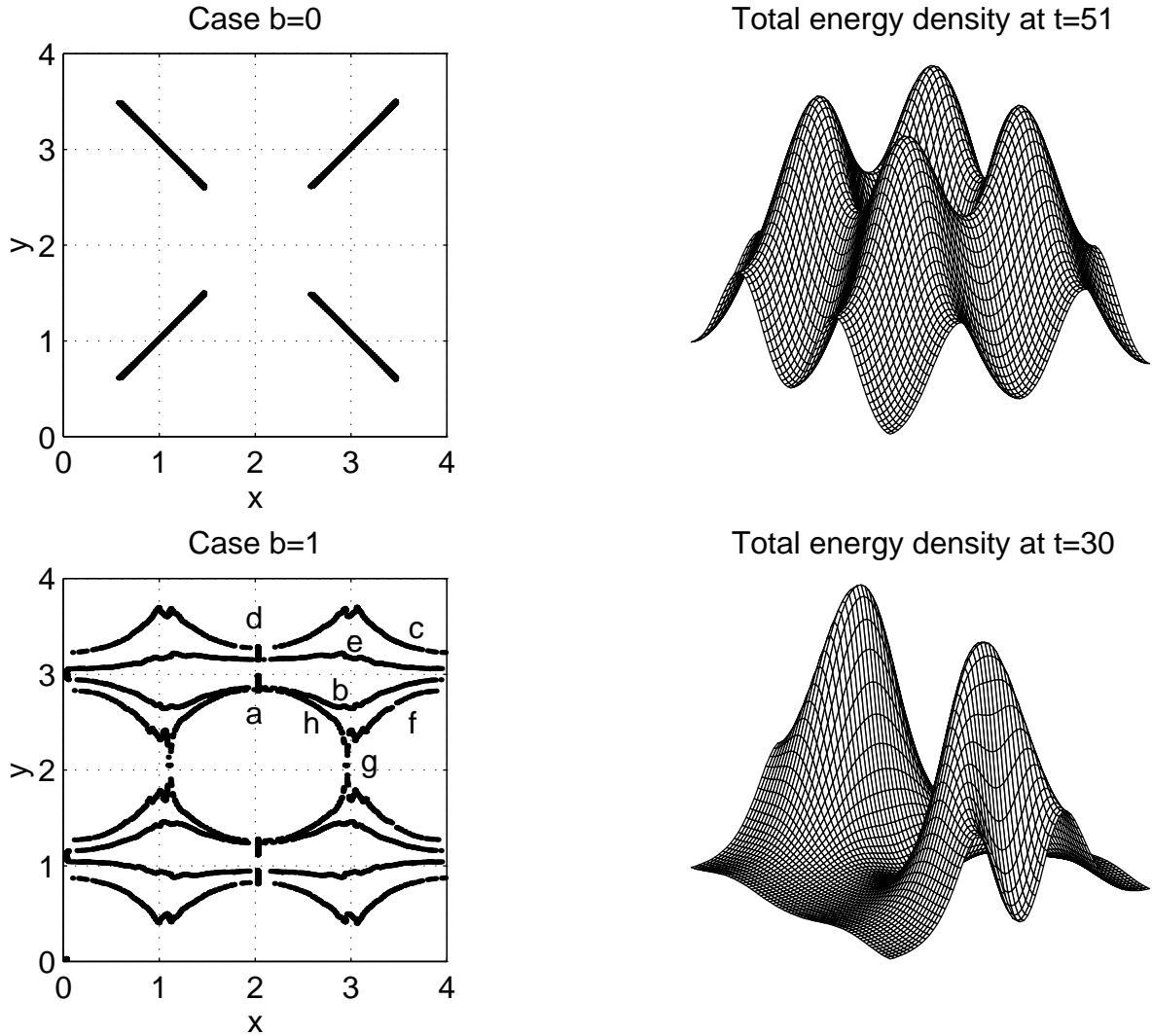}}
\caption{\underline{Above:} Trajectory of the skyrmions initially
on top of each other. They split up in four lumps heading to the corners
where they coalesce and break-off again, moving back to the centre of
the lattice, as in the illustration for $t$=51. \underline{Below:} The 
initially separated
skyrmions also divide each in two, but transit more complicated paths;
the signs $a-h$ refer to one of the `half-lumps'. The $t$=30
picture is shortly after the fractional progeny
have reunited at $d$ (and at its symmetrical point) and begun to travel
centrewards.}
\label{fig:pftwo}
\end{figure}

But in the stable, Skyrme situation, the lumps evolve in novel fashion.
Let us first consider the configuration when the extended entities are on
top of each other at $t=0$. As time elapses, the system splits in four
equal lumps, each progressing towards its nearest lattice corner. There
they meet and coalesce, for all corners are nothing but the same point.
Then the system splits up once more and the `fractional-skyrmions' make
their way back to the centre of the nett, in a cycle that repeats itself
indefinitely. The foregoing event is illustrated in the superior half of
figure \ref{fig:pftwo}, with the trajectory of the four energy peaks in
the $x-y$ plane. The accompanying 3-D picture captures the moment when the
skyrmion quartet, having concurred at the corners and split afresh, begin
to motion towards the centre of the network. Worthy of remark is that the
trajectory in question resembles the usual head-on collision course and
subsequent $90^\circ$ scattering of two solitons, in spite of ours being
energy chunks with no initial velocity. Also, we have thoroughly verified
that the topological number is 2 all along our numerical evolution, which
suggests that each `fractional soliton' carries a topological charge
of 1/2. 

Two $CP^1$-solitons on top of each other on $S_2$ have been numerically
studied in \cite{leese2}. Such initially coalescing objects (in the Skyrme
variation of the model) were found to move away from one another, two
evolving lumps in mutual repulsion. On $T_2$, however, we have observed a
qualitatively different phenomenon.

A more involved trajectory occurs for two initially well-separated skyrmions. 
In the bottom-left plot of figure \ref{fig:pftwo}, the labels $a-g$ indicate
the itinerary of one of the entities (call the corresponding symmetrical
points $a'-g'$). At $t=0$ a full lump is at point $a$ but it soon halves under
the numerical simulation.  One of its fractional offspring moves following
curve $b$, whereas its counterpart proceeds in the opposite sense. At $x=0=4$
they get back together into one full structure, which runs vertically up
before separating anew. One of these components cruises along $c$ and, at site
$d$, reunites back with its peer travelling from the left. Before dividing
itself according to curve $e$, the skyrmion is seen to shift towards the
centre, as one can tell from the small leg connecting curves $d$-$e$ (the full
lump started at $a'$ undergoes a similar process). The bottom-right
diagram of figure \ref{fig:pftwo} exhibits the skyrmions heading
centrewards from $d$ and $d'$. Thence our system continues through $f-g-h$
and returns to its $t=0$ coordinates.  Observe that at $g$ a half-soliton from
lump $a$ undistinguishably coalesces with a half-soliton from lump $a'$, so
actually we do not know which bit is ascending (descending) along $h$ ($h'$).
We terminated our simulations when a like cycle was about to commence, as
evidenced by the small vertical lines emerging from $a, a'$. 

We underline that our research has been constraint to systems with zero inital
speed. Important mathematical aspects of $CP^1$ solitons given by equation
(\ref{eq:w}) [the $O(3)$ case only] have been recently analysed in
\cite{martin}, where the dynamics of the solitons was carried out using the
geodesic approximation. The presence of four energy peaks rather than two is
therein discussed as well. Through our numerical approach, we are planning to
study ourselves the dynamics of soliton configurations defined by (\ref{eq:w}).

Finally, note that the study carried out in this article shows that both
repulsive and attractive interactions are operating between the skyrmions.

\newpage
\section{Concluding remarks}

The $CP^1$ model in (2+1) dimensions is variegated. More so its stable, skyrmionic
version on $T_2$, which in this work has been shown to possess qualitatively different
features as compared to the familiar model on the compactified plane. 

A particularity of our periodic solitons is that they have no analytic
representative of degree one, limitation dictated by their elliptic
nature. In (2+0) dimensions, the $CP^1$ model on $S_2$ is known to
have soliton solutions in all topological classes. 

Another peculiarity of the toroidal model, one herein discovered, is that the
properties of the skyrmions depend on the elliptic function used to define them.
  Thus, skyrmion fields expressed in terms of $\sigma(z)$, equation
(\ref{eq:wsigma}), evolve differently than those expressed through $\wp(z)$,
equation (\ref{eq:w}). In the former case, for $v_0=0$, the associated chunks of
energy stay still in their initial positions as time goes by. In the latter
case, the system splits up in four lumps that stroll the network, comeback
together, {\em etc.}, acted upon by repulsive-attrative forces. Consequently, it
would be interesting to analyse the solitons on $T_2$ by employing alternative
elliptic functions. Although in another context, a step in this direction has
already been taken in \cite{pms2}, where instantons on a torus are displayed in
terms of a Jacobian elliptic function. Let us remind that in $S_2$ no new traits
arise from casting two-soliton configurations in different ways, {\em e.g.}, \(
W= z^2,z^{-2}, \frac{(z-a)(z-b)}{(z-c)(z-d)} \). 

That the topological charge of our splitting system is very well
conserved (=2) as time progresses, invites speculation on whether 
each `fractional-soliton' carries a non-integer degree. 
Clearly, further research on this matter is required. 

Investigation on the appealing question of collisions between the
solitons (\ref{eq:w}) is currently under way. Note that already on
$T_2$, the fields (\ref{eq:wsigma}) have yielded -in the Skyrme case-
the outcome of always scattering off at 90$^{\circ}$ when impinged
with a non-zero initial speed \cite{non}. On $S_2$, unsimilarly,
the existence of a critical speed, below which the skyrmions scatter
at 180$^{\circ}$ to the initial direction of motion, has long been a
landmark of the planar model . 

\newpage 
\Large{\bf Acknowledgements} \\
\normalsize

I thank J. M. Speight for drawing my attention to $\wp$-defined
solitons, and W. J. Zakrzewski for enlightening conversations. The financial
support of {\em Universidad del Zulia} is acknowledged.



\begin{thebibliography}{99}
%
\bibitem{lit1} Eichenherr H. (1976) {\em Nucl. Phys.} {\bf B146} 215
%
\bibitem{lit2} Perelomov A. M. (1981) {\em Physica} {\bf 4D} 1
%
\bibitem{leese1} Leese R. A., Peyrard M. and
Zakrzewski W. J. (1990) {\em Nonlinearity} {\bf 3} 387
%
%
\bibitem{leese2} Leese R. A., Peyrard M. and Zakrzewski W. J.
(1990) {\em Nonlinearity} {\bf 3} 773
%
\bibitem{sky} Skyrme T. H. R. (1962) {\em Nucl. Phys.} {\bf 31} 556
%
\bibitem{pms1} Sutcliffe P. M. (1991) {\em Nonlinearity} {\bf 4} 1109
%
\bibitem{non} Cova R. J and Zakrzewski W. J. (1997) {\em Nonlinearity}
{\bf 10} 1305
%
\bibitem{goursat} Goursat E. (1916) 
{\em Functions of a complex variable} Dover Publications. Alternatively,
see: Erd\'{e}lyi A. {\em et. al.} (1953)
{\em Higher transcendental functions} vol II Mc Graw Hill
%
\bibitem{martin} Speight J. M. {\em Lump dynamics in the $CP^1$ model on the 
torus}, University of Texas at Austin preprint, Dept. of Math, Austin,
Texas, 78712, U.S.A 
%
\bibitem{pms2} Sutcliffe P. M. (1995) {\em Nonlinearity} {\bf 8}, 411
%
\end{thebibliography}
\end{document}